\documentclass[prd,twocolumn]{revtex4}
\usepackage{dcolumn}
\usepackage{multirow}
\usepackage{graphicx}
\usepackage{amssymb}
\usepackage{bm}
\usepackage{hyperref}
\usepackage{epstopdf}
\usepackage{color}
\usepackage{mathrsfs}
\usepackage{amsmath,amssymb,amsthm}
\usepackage{rotating}
\usepackage{sverb, longtable}
\usepackage{subfigure}

\usepackage[graphicx]{realboxes}
\usepackage{adjustbox}
\begin{document}

\title{Testing the first law of black hole mechanics with GW150914}
\author{Deng Wang}
\email{cstar@nao.cas.cn}
\affiliation{Instituto de F\'{i}sica Corpuscular (CSIC-Universitat de Val\`{e}ncia), E-46980 Paterna, Spain \\
	National Astronomical Observatories, Chinese Academy of Sciences, Beijing, 100012, China}
\begin{abstract}
Whether the first law of black hole mechanics is correct is an important question in black holes physics. Subjected to current limited gravitational wave events, we propose its weaker version that permits a relatively large perturbation to a black hole system and implement a simple test with the first event GW150914. Confronting the strain data with the theory, we obtain the constraint on the deviation parameter $\alpha=0.07\pm0.11$, which indicates that this weaker version is valid at the 68\% confidence level. This result implies that the first law of black hole mechanics may be correct.

\end{abstract}
\maketitle

\section{Introduction}
Black holes as the most intriguing compact objects predicted by general relativity (GR) \cite{Einstein1915} lie in the core of theoretical physics and astrophysics. Since initiated by Schwarzschild in 1916 \cite{Schwarzschild:1916uq}, the black hole physics has become a popular research field, especially, when one attempts to unify quantum mechanics with GR. The characteristic of a black hole is that it has an event horizon where a light ray can not escape from the tight binding of strong gravity. One of the most important aspects of black holes is the second law of black hole mechanics, i.e., Hawking's area theorem \cite{Hawking:1971tu}, which reveals that the horizon area of a black hole never decreases over time. It is a fundamental result from the cosmic censorship conjecture \cite{Penrose:1969pc} and GR. As products of GR, black holes can also be used for probing GR, although GR has performed successfully ranging from large scales to small scales \cite{Will:2014kxa,Ferreira:2019xrr}. It is natural that one can test GR by testing the correctness of the black hole area law.  

Up to now, there are two methods to test the black hole area law in the literature. The former is measuring the area variation between two progenitor black holes and the remnant black hole for a gravitational wave burst from a binary black hole merger \cite{Isi:2020tac}, while the latter is monitoring a specific black hole for a long time and measuring its area variation \cite{Wang:2022eod}. For the former case, the authors in \cite{Isi:2020tac} have confirmed the Hawking's area theorem by calculating the area variation between inspiral and ringdown phases for the first gravitational wave detection GW150914 from the LIGO and VIRGO collaboration \cite{LIGOScientific:2016aoc,LIGOScientific:2016lio,LIGOScientific:2019fpa,LIGOScientific:2020tif}.
For the latter case, using recent imaging observations of two supermassive black holes M87* \cite{EventHorizonTelescope:2019dse,EventHorizonTelescope:2019ggy,EventHorizonTelescope:2019pgp,EventHorizonTelescope:2019ths} and Sgr A* \cite{EventHorizonTelescope:2022wok,EventHorizonTelescope:2022xnr,EventHorizonTelescope:2022xqj} from the Event Horizon Telescope, the author in \cite{Wang:2022eod} has found that their horizon area variation fractions are compatible with the black hole area law at the $1\,\sigma$ confidence level.

Interestingly, the first law of black hole mechanics \cite{Bardeen:1973gs} is still not tested by using the gravitational wave data, but the second law has been fully explored through two different approaches. The first law of black hole mechanics states that an asymptotically flat stationary black hole spacetime can evolve to new stationary black hole spacetimes by considering perturbations (or infinite small variations). The first law reveals a deeply quantitative relation between different physical quantities of black holes, while the second law tells us the direction of time. As a consequence, the first law of black hole mechanics plays a crucial role in fundamental physics. In this study, we will test the first law of black hole mechanics by using the first binary black hole merger event GW150914.      

This study is outlined in the following manner. We describe the analysis methodology in the next section. In Section III, we display the numerical results. Discussions and conclusions are presented in the final section. We use the units $G=c=1$.  

\section{Methodology}
The first law of black hole mechanics, also known as Bekenstein-Smarr formula \cite{Bekenstein:1973ur,Smarr:1972kt}, reads as
\begin{equation}
\mathrm{d}M=\frac{\kappa}{8\pi}\mathrm{d}A+\Omega\mathrm{d}J+V\mathrm{d}Q, \label{1}
\end{equation}
where $M$, $\kappa$, $A$, $\Omega$, $J$, $V$ and $Q$ are the black hole mass, surface gravity, surface area, angular velocity, angular momentum, electronic potential and electronic charge, respectively. Since astrophysical black holes are almost free of charge, we neglect the last term in Eq.(\ref{1}). Furthermore, an accurate test of the first law of black hole mechanics requires a black hole with a small perturbation. However, current gravitational wave events do not have an ideal large mass ratio, and consequently, they can not provide a small enough perturbation. Therefore, we can not implement an accurate enough test now. But, we can carry out a rough test by rewriting Eq.(\ref{1}) as  
\begin{equation}
\mathrm{\Delta}M=(1-\alpha)\left(\frac{\kappa}{8\pi}\mathrm{\Delta}A+\Omega\mathrm{\Delta}J\right), \label{2}
\end{equation}
where the symbol $\Delta$ denotes the variation of a quantity at two times and $\alpha$ characterizes the deviation from this relation. For instance, $\mathrm{\Delta}M$ can be shown as $\mathrm{\Delta}M=M_2-M_1$ representing the mass difference at two time points $t_2$ and $t_1$. It is easy to find that Eq.(\ref{2}) is actually a weak version of the first law of black hole mechanics Eq.(\ref{1}). It allows the existence of relatively large perturbation of a black hole process, but may bring the loss of accuracy of the second law Eq.(\ref{1}). For current gravitational wave observations, one should at least test whether the weak version Eq.(\ref{2}) is valid.
$\kappa$, $A$, $\Omega$ and $J$ are expressed as
\begin{equation}
\kappa=\frac{\sqrt{1-\chi^2}}{2M(1+\sqrt{1-\chi^2})}, \label{3}
\end{equation}
\begin{equation}
A=8\pi M^2(1+\sqrt{1-\chi^2}), \label{4}
\end{equation}
\begin{equation}
\Omega=\frac{\chi}{2M(1+\sqrt{1-\chi^2})}, \label{5}
\end{equation}
\begin{equation}
J=\chi M^2, \label{6}
\end{equation}
where $\chi$ is the dimensionless spin magnitude.

\begin{figure}
	\centering
	\includegraphics[scale=0.5]{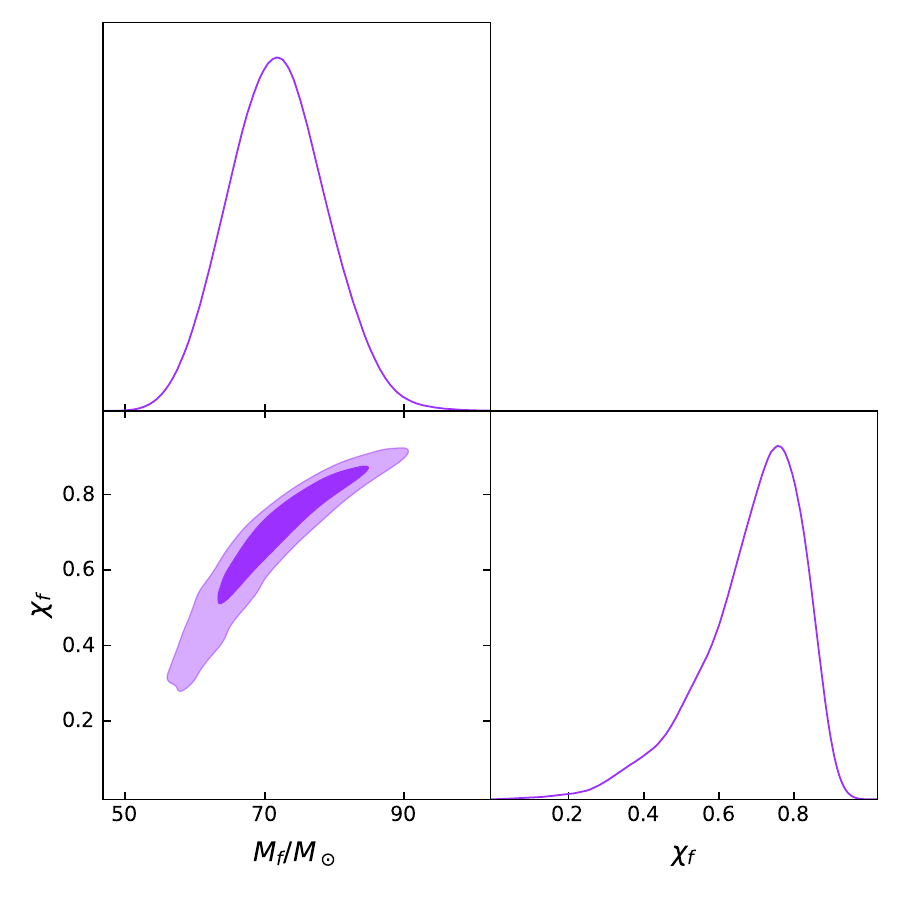}
	\caption{The marginalized posterior distributions of final mass and spin are shown by using the ringdown phase only data.}\label{f1}
\end{figure}

\begin{figure}
	\centering
	\includegraphics[scale=0.7]{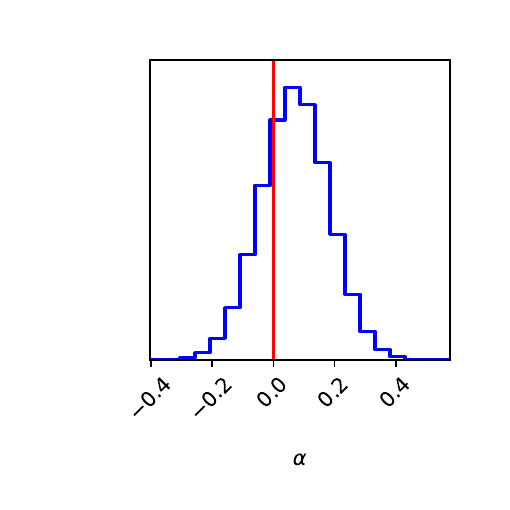}
	\caption{The 1-dimensional posterior distribution of the deviation parameter $\alpha$ is shown. The vertical red line denotes $\alpha=0$.}\label{f2}
\end{figure}

For a binary black hole coalescence, two primary black holes will experience three stages including inspiral, merger and ringdown and finally form the remnant large black hole. In light of this, we test the first law of black hole mechanics by analyzing independently the inspiral and ringdown parts of the signal in order to measure the possible variation in the total horizon area caused by the merger. To obtain the independent premerger and postmerger constraints, we divided the LIGO time-series data at the inferred peak of the characteristic strain into two separate parts. As a consequence, we can use the Bayesian analysis to estimate the black hole parameters of inspiral and ringdown phases. Note that the origin of data is 1126259462.423 s GPS in our analysis. 

For the inspiral phase, we employ the NRSUR7DQ4 template to fit the data up to the peak and estimate $M_1$, $M_2$, $\chi_1$ and $\chi_2$. Following Ref.\cite{Isi:2020tac}, We take
uniform priors for the total mass of the binary, mass ratio, spin magnitudes, luminosity volume, and the inclination, as well as an isotropic prior on the spin orientations. We also set the sky location as the values in Ref.\cite{Isi:2019aib}. For the ringdown phase, we take the black perturbation theory to derive the parameters of the final black hole by using the second part of GW150914 strain data. Specifically, we confronting the data with the simplest model without any overtone $3$ ms after the peak, namely the overtone number $N=0$ and the time delay after the inferred peak $\Delta t_0=3$ ms.

\section{Results}
By using the inspiral phase only data, we obtain the constraints on the progenitor black hole masses $M_1=39.1_{-7.3}^{+8.1}$, $M_2=32.7_{-7.8}^{+7.1}$ and their spins $\chi_1=0.37_{-0.08}^{+0.07}$ and $\chi_2=0.35_{-0.09}^{+0.07}$ at the 90\% confidence level. Furthermore, using the ringdown phase only data, we obtain the final mass $M_f=71.9_{-11.1}^{+11.5}$ and spin $\chi_f=0.69_{-0.31}^{+0.25}$ of the remnant balck hole based on simple black hole spectroscopy analysis at the 90\% confidence level (see Fig.\ref{f1}).   

To test the deviation from the first law of black hole mechanics, in the following analysis, we focus on studying the variations of physical quantities of the primary black hole with mass $M_1$. Inserting Eqs.(\ref{3}-\ref{6}) into Eq.(\ref{2}), one can derive the 1-dimensional probability distribution of the parameter $\alpha$. Furthermore, one can also derive its mean value and $1\,\sigma$ (68\%) uncertainty by the error propagation (see Fig.\ref{f2}). Finally, we obtain the $1\,\sigma$ constraint $\alpha=0.07\pm0.11$, which indicates that Eq.(2) holds at the 68\% confidence level and that the weaker version of the first law of black hole mechanics is valid. Furthermore, it also implies that the first law of black hole mechanics may be correct.    

\section{Discussions and conclusions}
In theoretical physics, the first law of black hole mechanics characterizes accurately the tiny evolution of a black hole system. Seen from a thermodynamical viewpoint, it describes the energy conservation of a black hole system. We wonder if this law is valid based on observations. In light of the first gravitational wave event GW150914, we propose a weaker version of this law and implement a simple test on it. We find that this weaker version that allows relatively large perturbations to a black hole is valid at the 68\% confidence level. This result suggests that the first law of black hole mechanics may be correct.  

{\bf Note added}. During the preparation of this manuscript, Ref.\cite{Wang:2023jah} appears. The authors use a black hole-neutron star candidate event, where the neutron star is regarded as a small perturbation to the black hole system, to test the first law. They also report a possible evidence that supports the validity of the first law of black hole mechanics.  

\section{Acknowledgments}
DW warmly thanks Yuan Sun for helpful discussions.

\end{document}